\documentclass[preprint,12pt]{elsarticle}
\usepackage{amsfonts}
\usepackage{amsmath}
\usepackage{amssymb}
\usepackage{bm}

\begin{document}

\begin{frontmatter}

\title{Superfluid phases of triplet pairing and rapid cooling of the neutron star in Cassiopeia A}
\author{Lev B. Leinson}
\address{Pushkov Institute of Terrestrial Magnetism, Ionosphere and Radiowave Propagation of the Russian 
Academy of Science (IZMIRAN),\\142190 Troitsk, Moscow, Russia}

\begin{abstract}
In a simple model it is demonstrated that the neutron star surface temperature evolution is sensitive to the phase state of the triplet superfluid condensate. A multicomponent triplet pairing of superfluid neutrons in the core of a neutron star with participation of several magnetic quantum numbers leads to neutrino energy losses exceeding the losses from the unicomponent pairing.  A phase transition of the neutron condensate into the multicomponent state triggers more rapid cooling of superfluid core in neutron stars. This makes it possible to simulate an anomalously rapid cooling of  neutron stars within the minimal cooling paradigm without employing any exotic scenarios suggested earlier for rapid cooling of isolated neutron star in Cassiopeia A.  
\end{abstract}
\begin{keyword}
Neutron star, Superfluidity, Neutrino radiation
\PACS 97.60.Jd, 26.60.Dd, 13.15.+g
\end{keyword}

\end{frontmatter}

\section{Introduction}

Studying thermal evolution of isolated neutron stars in X-rays is of a great
importance for better understanding the evolution of such objects and
provides a possibility to investigate their composition and structure (see
e.g., \cite{P04,P09,YP04}). The thermal X-ray radiation from the neutron
star (NS) at the center of the Cassiopeia A (Cas A) supernova remnant%
\footnote{%
The supernova remnant in Cassiopeia A contains a young ($\approx 330$ yr old 
\cite{F06}) neutron star which was discovered by Chandra satellite \cite%
{T99,H00} in 1999.} attracts much attention nowadays. A few years ago Heinke
\& Ho \cite{H09,H10} have analyzed Chandra observation data during 10 years
and reported an anomalous steady decline of the surface temperature, $T_{s}$%
. The authors have interpreted this data as a direct observation of Cas A NS
cooling, the phenomenon which has never been observed before for any
isolated NS.

We shall discuss later the current state of these observations, at the
moment we note that although the real cooling rate is under debate one can
not exclude that the Cas A NS cooling is extraordinarily fast. Even a $1\%$
decline of the cooling curve in 10 years would signal very fast cooling.
Such a rapid drop in surface temperature (if it occurs) is in conflict with
standard cooling scenarios based on the efficient modified Urca process. If
the NS in Cas A underwent standard cooling (through neutrino emission from
the core due to the modified Urca process) its surface temperature decline
in 10 years would be $0.2\%-0.3\%$ \cite{Y01,p06}.

The rapid decline but relatively high surface temperature (about $2.12\times
10^{6}$ K) require a dramatic change in the neutrino emission properties of
the NS. Some exotic scenarios of cooling have been suggested that employ
nonstandard assumptions on NS physics and evolution, involving softened pion
modes \cite{B12}, quarks \cite{s13,n13}, axions \cite{L14} or cooling after
an r-mode heating process \cite{y11}. An existence of softened pions or
quarks in the NS core depends mostly on the matter density but not on the
temperature. If this rapid cooling was constant from the birth of the NS,
the current temperature would have to be much smaller than is currently
measured.

It is reasonable to suggest \cite{P,St} that the cooling was initially slow
but greatly accelerated later. In this case the rapid temperature decline
could be naturally explainable in a frame of the minimal cooling paradigm 
\cite{P04,P09} that assumes that rapid cooling of the neutron star is
triggered by neutron superfluidity in the core. This scenario implies that
neutrons have recently become superfluid (in $^{3}$P$_{2}$ triplet-state) in
the NS core, triggering a huge neutrino flux from pair breaking and
formation (PBF) processes that accelerates the cooling \cite{P,St}, while
protons were already in a superconducting $^{1}$S$_{0}$ singlet-state with a
larger critical temperature. Although the above mechanism is consistent with
the commonly accepted cooling paradigm, the theoretical simulation has shown 
\cite{St,E13}, that the PBF processes in the neutron triplet condensate are
not enough effective to explain the rapid temperature decline. This has
stimulated the present work.

It is commonly believed \cite{T70,H70,Baldo,Elg} that the pair condensation
in the superdense neutron matter occurs into the $^{3}$P$_{2}$ state (with a
small admixture of $^{3}$F$_{2}$) with a preferred magnetic quantum number $%
m_{j}=0$. This model has been conventionally used for estimates of the PBF
neutrino energy losses in the minimal cooling scenarios.

Let us remind that, in the case of $^{3}$P$_{2}\left( m_{j}=0\right) $
pairing, the PBF $\bar{\nu}\nu $ emissivity is evaluated as \cite{L10} (we
use natural units, $\hbar =c=k_{B}=1$): 
\begin{equation}
Q(m_{j}=0)\simeq \frac{2}{5\pi ^{5}}G_{F}^{2}C_{\mathsf{A}}^{2}p_{F}M^{\ast
}T^{7}F\left( T/T_{c}\right) ~,  \label{Qnu}
\end{equation}%
where $G_{F}=1.166\times 10^{-5}$ GeV$^{-2}$ is the Fermi coupling constant, 
$C_{\mathsf{A}}$ is the axial-vector coupling constant of neutrons, $p_{F}$
is the Fermi momentum of neutrons, $M^{\ast }\equiv p_{F}/V_{F}$ is the
neutron effective mass; the function $F$ is given by 
\begin{equation}
F\left( T/T_{c}\right) =\int \frac{d\mathbf{n}}{4\pi }\frac{\Delta _{\mathbf{%
n}}^{2}}{T^{2}}\int_{0}^{\infty }dx\frac{z^{4}}{\left( \exp z+1\right) ^{2}},
\label{F0}
\end{equation}%
where $z=\sqrt{x^{2}+\Delta _{\mathbf{n}}^{2}/T^{2}}$, and the superfluid
energy gap, 
\begin{equation}
\Delta _{\mathbf{n}}\left( \theta ,T\right) =\sqrt{\frac{1}{2}\left( 1+3\cos
^{2}\theta \right) }\ \Delta \left( T\right) ,  \label{Dn}
\end{equation}%
is anisotropic. It depends on polar angle $\theta $ of the quasiparticle
momentum and temperature\footnote{%
Notice that our definition of the gap amplitude differs from the gap
definition used, in Ref. \cite{YKL} by the factor of $\sqrt{2}$.}.

In the present letter I argue that the enlarged neutrino energy losses can
be explained in terms of the conventional minimal cooling paradigm assuming
that the enhanced neutrino radiation can be a natural consequence of the
phase transition of the $^{3}$P$_{2}$ condensate into a multicomponent state.

Modern calculations \cite{Clark,Khodel} have shown that, besides the
one-component state with $m_{j}=0$, there are also multicomponent $^{3}$P$%
_{2}$ states involving several magnetic quantum numbers $m_{j}=0,\pm 1,\pm 2$
that compete in energy and represent various phases of the condensate in
equilibrium\footnote{%
Do not confuse with "angulons" which represent Goldstone bosons associated
with broken rotational symmetry in a $^{3}$P$_{2}\left( m_{j}=0\right) $
condensed neutron superfluid \cite{Bed13}. These collective excitations
represent small angular oscillations of the condensate. The complete set of
the oscillation modes of the $^{3}$P$_{2}\left( m_{j}=0\right) $ condensate
in the superfluid neutron liquid is analyzed in \cite{L10b}. Neutrino
emission due to decay of these collective oscillations produces a negligibly
small contribution into the NS cooling \cite{L13}.}. The general form of a
unitary $^{3}$P$_{2}$ state includes $m_{j}=0,\pm 1,\pm 2$, and the
superfluid energy gap can be defined by the relation \cite{L10a} 
\begin{equation}
D^{2}\left( \mathbf{n},\tau \right) =\mathbf{\bar{b}}^{2}\left( \mathbf{n}%
\right) \ \Delta ^{2}\left( \tau \right) ,  \label{D}
\end{equation}%
where $\tau \equiv T/T_{c}$ is the relative temperature; the (temperature
dependent) gap amplitude is of the form%
\begin{equation}
\Delta ^{2}=\Delta _{0}^{2}+2\Delta _{1}^{2}+2\Delta _{2}^{2}~,  \label{del}
\end{equation}%
and $\mathbf{\bar{b}}\left( \mathbf{n}\right) $ is a real vector normalized
by the condition 
\begin{equation}
\left\langle \bar{b}^{2}\left( \mathbf{n}\right) \right\rangle \equiv \left(
4\pi \right) ^{-1}\int \bar{b}^{2}\left( \mathbf{n}\right) d\mathbf{n}=1~.
\label{Norm}
\end{equation}%
Its angular dependence is represented by the unit vector $\mathbf{n=p}/p$
which defines the polar angles $\left( \theta ,\varphi \right) $ on the
Fermi surface:%
\begin{equation}
\mathbf{n=}\left( \sin \theta \cos \varphi ,\sin \theta \sin \varphi ,\cos
\theta \right) \equiv \left( n_{1},n_{2},n_{3}\right) .  \label{n}
\end{equation}%
The properly normalized vector $\mathbf{\bar{b}}$ can be written by
utilizing notation adopted in Refs. \cite{Clark,Khodel}, where $\lambda
_{1}\equiv \sqrt{6}\Delta _{1}/\Delta _{0}$ and $\lambda _{2}\equiv \sqrt{6}%
\Delta _{2}/\Delta _{0}$ : 
\begin{equation}
\mathbf{\bar{b}}=\sqrt{\frac{1}{2}}\frac{\Delta _{0}}{\Delta }\left( 
\begin{array}{ccc}
-n_{1}+n_{1}\lambda _{2}-n_{3}\lambda _{1}~, & -n_{2}-n_{2}\lambda _{2}~, & 
2n_{3}-n_{1}\lambda _{1}%
\end{array}%
\right) ~.  \label{burb}
\end{equation}

According to modern theories, there are several multicomponent states that
compete in energy depending on the temperature. Accordingly the phase
transitions can occur between these states when the temperature goes down.
The possible phase states of the $^{3}$PF$_{2}$ condensate are cataloged in
Ref. \cite{Clark}.

In Table 1 we have collected the nodeless states which are especially
interesting. Immediately below the critical temperature, the superfluid
condensate can appear in either the one-component phase $O_{0}$,
corresponding to $m_{j}=0$, or in one of the two two-component phases, $%
O_{\pm 3}$. These lowest-energy states are nearly degenerate. The higher
nearly degenerate group is composed of the phases $O_{1}$ and $O_{2}$.

The energy split between the two groups shrinks along with the temperature
decrease \cite{Clark} and can result in a phase transition at some
temperature\footnote{%
Authors predict the transition temperature $T\simeq 0.7T_{c}$ at $%
p_{F}\simeq 2.1$ fm$^{-1}$.} $T<T_{c}$, depending on the matter density. The
small difference in the gap amplitudes, $\sim 2\%$, inherent for various
phases of the condensate, is crucial for the phase transitions, but this
small inequality can be disregarded in evaluation of the neutrino energy
losses.

\begin{table}[tbp]
\caption{Various phases of the $^{3}$P$_{2}$ condensate and their relative
neutrino emissivity Z}%
\begin{tabular}{lcccccc}
&  &  &  &  &  &  \\ 
{phase} & {$\Delta_0 /\Delta$} & {$\lambda_1$} & {$\lambda_2$} & {$Z$} &  & 
\\ \hline\hline
$O_0$ & 1 & 0 & 0 & 1 &  &  \\ 
$O_{\pm 3}$ & $\frac{1}{2}$ & 0 & $\pm 3$ & $3.25$ &  &  \\ 
$O_{1} $ & $\frac{5}{\sqrt{14}\sqrt{17-3\sqrt{21}}}$ & $\frac{3}{5}\sqrt{%
2\left( 17-3\sqrt{21}\right) }$ & $\frac{3}{5}\left( \sqrt{21}-4\right)$ & $%
2.3528$ &  &  \\ 
$O_{2}$ & $\frac{5}{\sqrt{14}\sqrt{17+3\sqrt{21}}}$ & $\frac{3}{5}\sqrt{%
2\left( 17+3\sqrt{21}\right) }$ & $-\frac{3}{5}\left( \sqrt{21}+4\right)$ & $%
3.8258$ &  &  \\ 
&  &  &  &  &  & 
\end{tabular}%
\end{table}

\section{Neutrino emission from a multicomponent phase}

The neutrino emissivities of the multicomponent phase states have been
analyzed in Ref. \cite{L10a} in the approximation of averaged gap. The
calculation technique, developed in that work, allows us to derive a more
accurate expression taking into account the gap anisotropy. To this end we
have to use Eq (68) of Ref. \cite{L10a} and the polarization tensor, as
given just below Eq (65). Starting from these expressions we consider the
case of $\omega ^{2}>2\mathbf{\bar{b}}^{2}\Delta ^{2}$ which is fulfilled
for the PBF processes. Then after performing integrations over $d^{3}q$ one
can obtain the neutrino energy losses per unit volume and time in the $%
\Lambda $ state (we abbreviate the set of numbers $\Delta _{0}/\Delta
,\lambda _{1},\lambda _{2}$ as $\Lambda $). 
\begin{equation}
Q_{\Lambda }=\frac{2}{5\pi ^{5}}C_{A}^{2}G_{F}^{2}p_{F}M^{\ast
}T^{7}F_{\Lambda }\left( \tau \right) ~,  \label{Q}
\end{equation}%
where%
\begin{equation}
F_{\Lambda }\left( \tau \right) =\left( 4-3\frac{\Delta _{0}^{2}}{\Delta ^{2}%
}\right) y^{2}\int \frac{d\mathbf{n}}{4\pi }\bar{b}^{2}\left( \mathbf{n}%
\right) \int_{0}^{\infty }dx\frac{z^{4}}{\left( 1+\exp z\right) ^{2}}
\label{F}
\end{equation}%
with$\ z=\sqrt{x^{2}+\bar{b}^{2}\left( \mathbf{n}\right) y^{2}}$,~ $y\left(
\tau \right) =\Delta \left( T\right) /T$, and the function $\bar{b}%
^{2}\left( \mathbf{n}\right) $ is given by 
\begin{eqnarray}
\bar{b}^{2}\left( \mathbf{n}\right)  &=&\frac{1}{4}\frac{\Delta _{0}^{2}}{%
\Delta ^{2}}\left[ 2+\lambda _{1}^{2}+2\lambda _{2}^{2}+\left( 6+\lambda
_{1}^{2}-2\lambda _{2}^{2}\right) \cos ^{2}\theta \right.   \notag \\
&&\left. -2\lambda _{1}\left( 1+\lambda _{2}\right) \sin 2\theta \,\cos
\varphi +\left( \lambda _{1}^{2}-4\lambda _{2}\right) \sin ^{2}\theta \,\cos
2\varphi \right]   \label{b2}
\end{eqnarray}%
At $\lambda _{1}=\lambda _{2}=0$ and $\Delta =\Delta _{0}$ the expression (%
\ref{Q}) recovers Eq. (\ref{Qnu}).

For numerical evaluation of the neutrino losses, as given in Eq. (\ref{Q}),
it is necessary to know the function $y\left( \tau \right) =\Delta \left(
T\right) /T$, which in general is to be found with the aid of gap equations.
However, as mentioned above, the difference in the gap amplitudes for
various phases can be neglected in evaluation of the neutrino energy losses.
This substantially simplifies the problem because for the case $m_{j}=0$ the
function is well investigated\footnote{%
We use the simple fit $\sqrt{2}\mathsf{v}_{B}\left( \tau \right) $ suggested
in Ref. \cite{YKL}.}.

\section{Modeling of the cooling process}

To get an idea of how the phase state of the superfluid condensate can
influence the NS surface temperature let us consider a simple model of
cooling of the superfluid neutron core enclosed in a thin envelope.

We assume that the bulk matter consists mostly of $^{3}$P$_{2}$ superfluid
neutrons. The neutrino emission due to $^{1}$S$_{0}$ proton pairing is
strongly suppressed in the non-relativistic system \cite{FRS76,LP06}, but
the energy gap arising in the quasiparticle spectrum below the condensation
temperature suppresses the most mechanisms of neutrino emission which are
efficient in the normal (nonsuperfluid) nucleon matter ($\nu \bar{\nu}$
bremsstrahlung, modified Urca processes etc.) \cite{YLS}. As was found in
Ref. \cite{St,P} this scenario puts stringent constraints on the temperature
for the onset of neutron superfluidity in the Cas A NS. Namely, the
transition temperature dependence on the density should have a wide peak
with maximum $T_{c}(\rho )\approx (5-8)\times 10^{8}$~K.

In the temperature range which we are interested in, the thermal luminosity
of the surface is negligible in comparison to the neutrino luminosity of PBF
processes in the NS core. In this case the equation of global thermal
balance \cite{gs80} reduces to 
\begin{equation}
C(\widetilde{T})\,{\frac{d\widetilde{T}}{dt}}=-L(\widetilde{T}).  \label{aa}
\end{equation}%
Here $L(\widetilde{T})$ is the total PBF luminosity of the star (redshifted
to a distant observer), while $C(\widetilde{T})$ is the stellar heat
capacity. These quantities are given by (see details in Ref. \cite{Y}): 
\begin{eqnarray}
L(\widetilde{T}) &=&\int dV\,Q_{\Lambda }(T,\rho )\exp (2\Phi (r)),
\label{eq:Lnu} \\
C(\widetilde{T}) &=&\int dV\,C_{V}(T,\rho ),  \label{C}
\end{eqnarray}%
where $C_{V}(T,\rho )$ is the specific heat capacity, 
\begin{equation*}
dV=4\pi r^{2}\left( 1-\frac{2Gm(r)}{r}\right) ^{-1/2}dr,
\end{equation*}%
where $G$ stands for gravitation constant, $m\left( r\right) $ is the
gravitational mass enclosed within radius $r$, and $\Phi (r)$ is the metric
function that determines gravitational redshift. A thermally relaxed star
has an isothermal interior which extends from the center to the heat
blanketing envelope. Following \cite{gs80} we have assumed that the
isothermal region is restricted by the condition $\rho >\rho \left( r_{%
\mathsf{b}}\right) =10^{10}$ \textrm{g cm}$^{-3}$. Taking into account the
effects of General Relativity (e.g., \cite{thorne77}), isothermality at $%
r<r_{\mathsf{b}}$ means spatially constant redshifted internal temperature $%
\widetilde{T}(t)$, while the local internal temperature 
\begin{equation}
T(r,t)=\widetilde{T}(t)\exp \left( -\Phi (r)\right) ,  \label{bb}
\end{equation}%
depends on radial coordinate $r$. Generally, the redshift factor has to be
calculated using the Tolman-Oppenheimer-Volkoff equation. In vacuum, outside
the star and at the stellar surface this factor is of the form 
\begin{equation}
\exp \Phi (r)=\left( 1-\frac{2Gm(r)}{r}\right) ^{1/2}.  \label{ex}
\end{equation}%
For simplicity we shall use this expression in the crust of the star, as a
model.

The main temperature gradient is formed in the thermally insulating outer
envelope at $r>r_{\mathsf{b}}$. Since the envelope is thin one can set $r_{%
\mathsf{b}}\simeq R$ and $m\left( r_{\mathsf{b}}\right) \simeq M$, where $R$
and $M$ are the radius and mass of the NS, respectively. Then the
temperature $T_{\mathsf{b}}=T\left( r_{\mathsf{b}}\right) $ at the bottom of
the thermally insulating envelope of the star can be written as%
\begin{equation}
T_{\mathsf{b}}=\left( 1-\frac{R_{g}}{R}\right) ^{-1/2}\widetilde{T},
\label{TbT}
\end{equation}%
where 
\begin{equation}
R_{g}\equiv 2GM\simeq 2.953\frac{M}{M_{\odot }}~\mathrm{km}  \label{x}
\end{equation}%
is the Schwarzschild radius.

One can convert the internal $T_{\mathsf{b}}$ to the observed effective
surface temperature $T_{\mathsf{s}}$ using the simple analytical
relationship found by Gundmundsson, Pethick \& Epstein \cite{GPE}: 
\begin{equation}
T_{\mathsf{s}}/10^{6}\mathrm{K}\simeq 0.87g_{s14}^{1/4}(T_{\mathsf{b}}/10^{8}%
\mathrm{K})^{0.55}.  \label{TeTb}
\end{equation}%
Here $g_{s14}=g_{s}/10^{14}\mathrm{cm~s}^{-2}$ where 
\begin{equation}
g_{s}=\frac{GM}{R^{2}\sqrt{1-R_{g}/R}}\simeq \frac{1.328\times 10^{14}}{%
\sqrt{1-R_{g}/R}}\frac{M/M_{\odot }}{R_{6}^{2}}~\mathrm{cm~s}^{-2},
\label{gs}
\end{equation}%
with $R_{6}\equiv R/\left( 10^{6}\mathrm{cm}\right) $, is the acceleration
of gravity as measured at the surface.

Given the strong dependence of the PBF processes on the temperature $T$ and
density $\rho $, the overall effect of emission of neutrino pairs can only
be assessed by complete calculations of the neutron star cooling which are
beyond the scope of this paper. We do not aim to carry out exact
calculations. Our goal is to demonstrate that the NS cooling rate
substantially depends on the phase state of the $^{3}$P$_{2}$ condensate of
superfluid neutrons. A rough estimate can be made in a simplified model,
where both the superfluid transition temperature, $T_{c}$, and the real
temperature, $T=T_{\mathsf{core}}$, are constant over the core.

In the temperature range of our interest, the specific heat is governed by
the neutron component (the contribution of electrons and strongly superfluid
protons is negligibly small) and can be described as 
\begin{equation}
C\simeq \frac{1}{3}T_{\mathsf{core}}R_{B}(T_{\mathsf{core}}/T_{c})\int
dVp_{F}M^{\ast },  \label{cc}
\end{equation}%
where $R_{B}(\tau )$ is the superfluid reduction factor, as given in Eq.
(18) of Ref. \cite{YLS}.

Making use of Eq. (\ref{Q}) we obtain the PBF luminosity in the form 
\begin{equation}
L=\frac{2}{5\pi ^{5}}G_{F}^{2}C_{\mathsf{A}}^{2}T_{\mathsf{core}%
}^{7}F_{\Lambda }(T_{\mathsf{core}}/T_{c})\int dVp_{F}M^{\ast }e^{2\Phi (r)},
\label{dd}
\end{equation}%
where $F_{\Lambda }(\tau )$ is given by Eq. (\ref{F}).

Insertion of Eqs. (\ref{bb}), (\ref{cc}) and (\ref{dd}) into Eq. (\ref{aa})
allows us to obtain the following equation for the non-redshifted
temperature $T(r_{\mathsf{core}},t)\equiv T_{\mathsf{core}}(t)$ at the edge
of the core, at $r=r_{\mathsf{core}}$: 
\begin{equation}
\frac{dT_{\mathsf{core}}}{dt}=-\frac{3\alpha }{R_{B}\left( T_{\mathsf{core}%
}/T_{c}\right) }\frac{2}{5\pi ^{5}}G_{F}^{2}C_{\mathsf{A}}^{2}T_{\mathsf{core%
}}^{6}F_{\Lambda }\left( T_{\mathsf{core}}/T_{c}\right) .  \label{Tbeq}
\end{equation}%
Here the constant $\alpha \equiv \alpha (r_{\mathsf{core}})$ is defined as 
\begin{equation}
\alpha \equiv \frac{\int dVp_{F}M^{\ast }e^{2\Phi \left( r\right) }}{\exp
\Phi \left( r_{\mathsf{core}}\right) \int dVp_{F}M^{\ast }},  \label{alpha}
\end{equation}%
where the integration is over the core volume, $r\leq r_{\mathsf{core}}$.

In Eq. (\ref{Tbeq}) $T_{\mathsf{core}}$ is the real temperature in the core,
particularly, at the crust-core interface which corresponds to the density
of about $1.5\times 10^{14}$ \textrm{g/cm}$^{3}$ at $r=r_{\mathsf{core}}$.
One can convert it to the redshifted internal temperature $\widetilde{T}(t)$%
~as%
\begin{equation}
\widetilde{T}=\left( 1-\frac{2Gm\left( r_{_{\mathsf{core}}}\right) }{r_{_{%
\mathsf{core}}}}\right) ^{1/2}T_{\mathsf{core}}\simeq \left( 1-\frac{R_{g}}{%
r_{_{\mathsf{core}}}}\right) ^{1/2}T_{\mathsf{core}}.  \label{Tt}
\end{equation}%
When obtaining the second equality we have neglected the mass of the crust
which is small ($\sim 1\%$) in comparison with the mass of the core \cite%
{PR95}. This allows us to set $m\left( r_{_{\mathsf{core}}}\right) \simeq M$%
. \ 

From Eqs. (\ref{TbT}) and (\ref{Tt}) one can find the temperature at the
bottom of the thermally insulating envelope%
\begin{equation}
T_{\mathsf{b}}=\left( 1-\frac{R_{g}}{R}\right) ^{-1/2}\left( 1-\frac{R_{g}}{%
r_{_{\mathsf{core}}}}\right) ^{1/2}T_{\mathsf{core}}.  \label{TbTc}
\end{equation}%
Insertion of this expression into Eq. (\ref{TeTb}) allows one to find the
observed (non-redshifted) surface temperature $T_{\mathsf{s}}$: 
\begin{equation}
T_{\mathsf{s}}/10^{6}\mathrm{K}\simeq 0.87g_{s14}^{1/4}\left( \frac{%
1-R_{g}/r_{_{\mathsf{core}}}}{1-R_{g}/R}\right) ^{\frac{0.55}{2}}(T_{\mathsf{%
core}}/10^{8}\mathrm{K})^{0.55}.  \label{Ts}
\end{equation}%
$\,$Assuming that the crust thickness is about $0.1R$ \cite{PR95} one can
set $r_{\mathsf{core}}\simeq 0.9R$.

We adopt $R=10.3~\mathrm{km}$ and $M=1.65M_{\odot }$. In this case 
\begin{equation}
0.87g_{s14}^{1/4}\left( \frac{1-R_{g}/r_{_{\mathsf{core}}}}{1-R_{g}/R}%
\right) ^{\frac{0.55}{2}}\simeq 1.\,\allowbreak 098,  \label{factor}
\end{equation}%
which yields%
\begin{equation}
T_{\mathsf{s}}/10^{6}\mathrm{K}\simeq 1.\,\allowbreak 098(T_{\mathsf{core}%
}/10^{8}\mathrm{K})^{0.55}.  \label{Tb}
\end{equation}%
Thus our simulation of the NS cooling is reduced to numerical solving of
Eqs. (\ref{Tbeq}) and (\ref{Tb}).

\section{Simulation results}

In Fig. 1 we demonstrate the cooling curves of the superfluid neutron star
with a constant $T_{c}$ over the core. The curves obtained for the
superfluid phases listed in Table 1 and are labeled respectively.

\begin{figure}[h]
\includegraphics{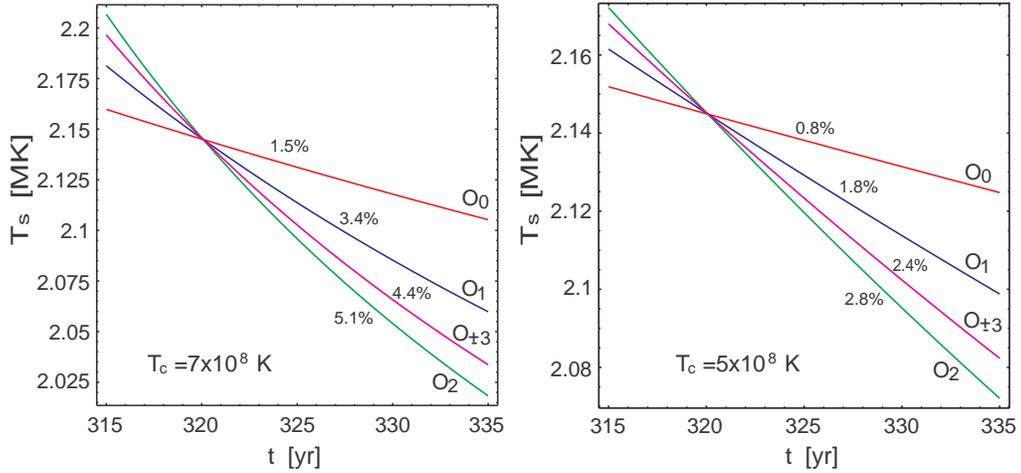}
\caption{(Color on line) \textit{Left:} Cooling curves for Cas A NS which
has a superfluid neutron core and a low-mass heat blanketing envelope. $%
T_{c}=7\times 10^{8}$~K is taken constant over the core. Four curves
correspond to different phases of triplet pairing. $O_{0}$ is the cooling
curve of the one-component phase $m_{j}=0$. The remaining curves correspond
to the $O_{1}$, $O_{2}$, and $O_{\pm 3}$ phases. Calculated temperature
declines over 10 years are given near the curves (in percent). \textit{Right:%
} Same but with $T_{c}=5\times 10^{8}$~K. }
\label{fig:fig1}
\end{figure}

The case $O_{0}$ corresponds to the one-component state of the neutron
superfluid with $m_{j}=0$. The remaining three curves correspond to the
phases $O_{1}$, $O_{2}$ and $O_{\pm 3}$. Two panels of Figure 1 demonstrate
the corresponding simulated cooling curves for the cases of $T_{c}=7\times
10^{8}$ K and $T_{c}=5\times 10^{8}$ K. We show the cooling curves over a
period of about 25 years including 10 years of observations. Note that we
show the non-redshifted effective surface temperature. Calculated
temperature declines over 10 years are given near the curves (in percent).
As it is seen from these curves a satisfactory agreement with observable
temperature declines can be easily obtained by a proper choice of the phase
state of $^{3}$P$_{2}$ condensate and adjusting the parameters of
superfluidity.

Certainly the approximation of the constant superfluid transition
temperature over the neutron star core is too crude, and simulations with
realistic $T_{c}\left( \rho \right) $ profile can be more persuasive. Such a
numerical simulation is beyond the scope of this work. Although it is
necessary to note that similar simulations were done in \cite{E13}, where
five phenomenological $T_{c}\left( \rho \right) $ profiles over the NS core
were considered, but the free parameter was used for artificial increase of
the PBF neutrino emissivity from the $^{3}$P$_{2}\left( m_{j}=0\right) $
pairing. These more realistic calculations are in agreement with our
qualitative estimates. Our primary goal is to clarify the possible origin
for the increased neutrino losses.

One can make a simple estimate of the relative efficiency of PBF processes
for various phases of the superfluid neutron matter. To this end we can
evaluate Eq. (\ref{Q}) in the approximation of averaged gap that reduces to
the replacement $\bar{b}^{2}\rightarrow \left\langle \bar{b}%
^{2}\right\rangle =1$. We then recover the result obtained in Eq. (74) of
Ref. \cite{L10a}):%
\begin{equation}
\bar{Q}_{\Lambda }\simeq Z\left( \Lambda \right) \bar{Q}(m_{j}=0)~,
\label{Qm}
\end{equation}%
where $\bar{Q}(m_{j}=0)$ is given by Eq. (\ref{Qnu}) but with a replacement $%
\Delta _{\mathbf{n}}^{2}\rightarrow \Delta ^{2}$, and%
\begin{equation}
Z\left( \Lambda \right) =\left( 4-3\frac{\Delta _{0}^{2}}{\Delta ^{2}}%
\right) ~,  \label{r}
\end{equation}%
These factors representing the relative efficiency of PBF processes for
various phases of the $^{3}$P$_{2}$ superfluid neutron matter are shown in
Table 1.

\section{Discussion and conclusion}

Our simple analytic expression (\ref{Q}) for the PBF neutrino emissivity
from the multicomponent phases of the $^{3}$P$_{2}$ superfluid neutron
liquid shows that the PBF neutrino losses from the multicomponent condensate
can be a few times larger than the corresponding neutrino losses from the
one-component condensate with $m_{j}=0$.

We have employed Eq. (\ref{Q}) for a simple cooling model of a superfluid
neutron core enclosed in a thin envelope assuming that the superfluid
transition temperature $T_{c}$ is constant over the core. In this simple
model we have demonstrated that the NS surface temperature is sensitive to
the phase state of the superfluid condensate of neutrons, and this allows
one to qualitatively explain the anomalously rapid cooling of the Cas A NS
(if it occurs). In other words, we have demonstrated the principal
possibility of simulations of rapid cooling in frame of the minimal cooling
paradigm without any artificial change of the PBF neutrino emissivity from
the $^{3}$P$_{2}(m_{j}=0)$ pairing, as was suggested in Refs. \cite{St,E13}.
In a realistic case the superfluid transition temperature $T_{c}$ as well as
the phase state of the condensate are dependent on the matter density and
therefore the phase state of the superfluid liquid can vary along with the
distance from the core center. However, the qualitative effects will not be
modified by the inclusion of more realistic physics. All the effects
discussed above make it possible to explain an anomalously rapid cooling of
NSs in many details.

The involving relevance of the multicomponent condensation of neutrons into
simulation of the Cas A NS cooling depends on its actual cooling rate which
is controversial at the moment. Heinke \& Ho \cite{H09,H10} have analyzed
the archival data from the Chandra X-ray Observatory ACIS-S detector in
Graded mode between 2000 and 2009 and reported a steady decline of the
surface temperature, $T_{s}$, by about $4\%$. New observational work on Cas
A has shown, however, that the above mentioned rapid cooling of the Cas A NS
is not so evident due to systematic uncertainties inherent in the
observations and associated with calibration problems of Chandra detectors 
\cite{E13,P13}.

Elshamouty et al. \cite{E13} compared the results from all the Chandra
detectors and found the weighted mean of the temperature decline rate of $%
2.9\pm 0.5_{\mathsf{stat}}\pm 1_{\mathsf{sys}}\%$ over 10 years of
observations using the data of all detectors, and a weaker decline of $%
1.4\pm 0.6_{\mathsf{stat}}\pm 1_{\mathsf{sys}}\%$ excluding the data from
the ASIS-S detector in the graded mode which suffers from the grade
migration.

In contrast, Posselt et al. \cite{P13} do not confirm the existence of
statistically significant temperature decline and attribute the observed
effect to the degradation of the Chandra ASIS-S detector in soft channels.
The authors state that the previously reported rapid cooling of the Cas A NS
is likely a systematic artifact, and they cannot exclude the standard slow
cooling for this NS. Their results (2006-2012) are consistent with no
temperature decline at all, or a smaller temperature decline than that
reported before although the involved uncertainties are too large to firmly
exclude the previously reported fast cooling.

Further observations are necessary to assess the rate of temperature drop
with higher accuracy. Let us notice, however, that the discussed problem of
the multicomponent condensation of neutrons can be of interest not only to
the Cas A NS cooling but can be relevant also for other superfluid NSs.


\begin{thebibliography}{99}
\bibitem{P04} D. Page, J.M. Lattimer, M. Prakash, and A.W. Steiner,
Astrophys. J. Supp. 155 (2004) 623.

\bibitem{P09} D. Page, J.M. Lattimer, M. Prakash, and A.W. Steiner,
Astrophys. J. 707 (2009) 1131.

\bibitem{YP04} D. G. Yakovlev, \& C. J. Pethick, ARA\&A, 42 (2004) 169.

\bibitem{F06} R. A. Fesen et al., Astrophys. J. 645 (2006) 283.

\bibitem{T99} H. Tananbaum, IAU Circ., 7246 (1999) 1.

\bibitem{H00} J. P. Hughes et al., Astrophys. J. Lett. 528 (2000) L109.

\bibitem{H09} W.C.G. Ho, C.O. Heinke, Nature 462 (2009) 71.

\bibitem{H10} Heinke C.O., Ho W.C.G., Astrophys. J 719 (2010) L167.

\bibitem{Y01} D. G. Yakovlev et al., Phys. Rept. 354 (2001) 1.

\bibitem{p06} D. Page, U. Geppert, and F. Weber, Nucl. Phys. A {777} (2006)
497.

\bibitem{B12} D. Blaschke, H. Grigorian, D.N. Voskresensky, and F. Weber,
Phys. Rev. C 85 (2012) 022802.

\bibitem{s13} A. Sedrakian, Astron.Astrophys. 555 (2013) L10

\bibitem{n13} T. Noda, M.-a. Hashimoto, N. Yasutake, et al., Astrophys. J.
765 ( 2013) 1.

\bibitem{L14} L.B. Leinson, J. Cosmol. Astropart. Phys. 08 (2014) 031.

\bibitem{y11} S.-H. Yang, C.-M. Pi, \& X.-P. Zheng, Astrophys. J. 735 (2011)
L29.

\bibitem{P} D. Page, M. Prakash, J.M. Lattimer and A.W. Steiner, Phys. Rev.
Lett. 106 (2011) 081101.

\bibitem{St} P.S. Shternin, D.G. Yakovlev, C.O. Heinke, W.C.G. Ho, D.J.
Patnaude, Mon. Not. R. Astron. Soc. 412 (2011) L108.

\bibitem{E13} K.G.Elshamouty, C.O. Heike, G.R. Sivakoff, W.C.G. Ho, P.S.
Shternin, D.G. Yakovlev, D.J. Patnaude, \& L. David, Astrophys. J. 777
(2013) 22.

\bibitem{T70} R. Tamagaki, Prog. Theor. Phys. 44 (1970) 905.

\bibitem{H70} M. Hoffberg, A.E. Glassgold, R.W. Richardson, and M. Ruderman,
Phys. Rev. Lett. 24 (1970) 775.

\bibitem{Baldo} M. Baldo, J. Cugnon, A. Lejeune and U. Lombardo, Nucl. Phys.
A 536 (1992) 349.

\bibitem{Elg} \O . Elgar{\o }y, L. Engvik, M. Hjorth-Jensen, E. Osnes, Nucl.
Phys. A 607 (1996) 425.

\bibitem{L10} L.B. Leinson, Phys. Rev. C 81 (2010) 025501.

\bibitem{YKL} D.G. Yakovlev, A.D. Kaminker and K.P. Levenfish, Astron.
Astrophys. 343 (1999) 650.

\bibitem{Clark} V.A. Khodel, J.W. Clark, and M.V. Zverev, Phys. Rev. Lett.
87 (2001) 031103.

\bibitem{Khodel} M.V. Zverev, J.W. Clark, V.A. Khodel, Nucl. Phys. A 720
(2003) 20.

\bibitem{Bed13} P. F. Bedaque, and A.N. Nicholson, Phys. Rev. C87 (2013)
055807, Erratum-ibid. C89 (2014) 029902.

\bibitem{L10b} L.B. Leinson, Phys. Rev. C 85 (2012) 065502.

\bibitem{L13} L.B. Leinson, Phys. Rev. C 87 (2013) 025501.

\bibitem{L10a} L.B. Leinson, Phys. Rev. C 82 (2010) 065503.

\bibitem{FRS76} E. Flowers, M. Ruderman, P. Sutherland, Astrophys. J. 205
(1976) 541.

\bibitem{LP06} L.B. Leinson and A. P\'{e}rez, Phys. Lett. B 638 (2006) 114.

\bibitem{YLS} D.G. Yakovlev, K.P. Levenfish, Yu.A. Shibanov, Phys. Usp. 42
(1999) 737.

\bibitem{gs80} G. Glen, P. Sutherland, Astrophys. J. 239 (1980) 671.

\bibitem{Y} D.G. Yakovlev, W.C.G. Ho, P.S. Shternin, C.O. Heinke and A.Y.
Potekhin, Mon. Not. R. Astron. Soc. 411 (2011) 1977.

\bibitem{thorne77} K.S. Thorne, Astrophys. J. 212 (1977) 825.

\bibitem{PR95} C.J. Pethick and D.G. Ravenhall, Annu. Rev. Nucl. Part. Sci.
45 (1995) 429.

\bibitem{GPE} E.H. Gudmundsson, C.J. Pethick, R.I. Epstein, Astrophys. J.
259 (1982) L19.

\bibitem{P13} B. Posselt, G.G. Pavlov, V. Suleimanov, O. Kargaltsev,
Astrophys. J. 779 (2013) 186.
\end{thebibliography}
\end{document}